\documentclass[
aip,
reprint,
amsmath,amssymb,
 amsmath,amssymb,
 aps,
]{revtex4-2}
\usepackage[T1]{fontenc}
\usepackage{times}
\usepackage{graphicx}
\usepackage{amsfonts, amsmath,amssymb, bm, graphicx}
\usepackage{siunitx}
\usepackage{ulem}
\usepackage{amssymb}
\usepackage[english]{babel}
\usepackage{lipsum}
\usepackage{mathtools}
\usepackage{dcolumn} 
\usepackage{bm} 
\usepackage[english]{babel}
\usepackage{amssymb,amsmath}
\usepackage{hyperref}
\usepackage{color}
\usepackage{bm}
\usepackage{xcolor}
\usepackage{siunitx}

\begin{document}

\preprint{APS/123-QED}

\title{Equilibrium Magnetic Properties in Magnetic Nanoscrews}

\author{Victoria Acosta-Pareja}
\affiliation{Universidad Cat\'olica del Norte, Avenida Angamos 0610, Antofagasta, Chile}

\author{ Valeria M. A. Salinas}
\affiliation{Universidad Cat\'olica del Norte, Avenida Angamos 0610, Antofagasta, Chile}

\author{Omar J. Suarez}
\affiliation{Departamento de F\'{i}sica, Universidad de Sucre,  Sincelejo, Colombia}

\author{Attila K\'akay}
\affiliation{Helmholtz-Zentrum Dresden-Rossendorf (HZDR), Dresden, Germany}

\author{Jorge A. Ot\'alora}
\email{jorge.otalora@ucn.cl}
\affiliation{Universidad Cat\'olica del Norte, Avenida Angamos 0610, Antofagasta, Chile}

\date{\today}




\date{\today}

\begin{abstract}
We investigate the equilibrium magnetization in ferromagnetic nanoscrews (NSw) using micromagnetic simulations. These systems consist of elongated three-dimensional magnetic membranes with helicoidal geometry, combining curvature, torsion ($\mathrm{w}$), and eccentricity ($\epsilon$) along their length. We focus on the influence of these geometric parameters, together with membrane thickness and inner diameter, on remanent states and coercive fields.
Our results, obtained over a broad range of eccentricities and torsions, reveal bistable magnetic behavior, with vortex-domain-wall propagation during magnetization reversal. We identify four degenerate configurations of a remarkably stable mixed remanent state. The coercive field is found to increase with eccentricity for structures with a major axis (larger inner diameter) approximately 30\% larger than the minor axis (smaller inner diameter), while remaining largely insensitive to variations in torsion.
These findings are interpreted in terms of geometry-induced modifications of surface magnetostatic charges on the membrane mantle. Overall, our results demonstrate that nanoscrews exhibit robust bistability under systematic geometric deformation, together with enhanced coercivity, highlighting their potential for applications in three-dimensional nanomagnetism.
\end{abstract}

\keywords{Nanoscrew, Equilibrium states}
\maketitle


\section{\label{sec:level1}INTRODUCTION}

In recent decades, the field of nanomagnetism has undergone a significant transition from two-dimensional planar systems towards three-dimensional (3D) curvilinear magnetic nanostructures of increasing complexity \cite{Three, li2023magnetic, APLMATPACHECO2020, chumak2022advances, Gubbiotti_2025}. Beyond miniaturization, this development has revealed that geometry acts as an active control parameter capable of modifying the balance between fundamental magnetic interactions. In curved nanostructures, curvature and torsion renormalize the exchange and dipolar energies, thereby influencing their competition and leading to equilibrium configurations that differ substantially from those in planar systems. A variety of geometries, including nanowires, nanotubes, nanorings, M\"obius structures, nanohelices, and rolled membranes, have demonstrated that geometric confinement and symmetry breaking can strongly influence magnetic stability and reversal mechanisms \cite{Fedorov2024NatComm, Streubel2016JPD, chesun, Yershov2015PRB, Yershov2016PRB, vale, Sheka2015PRB, Mobius, Omar, Han2009AdvMater31, ucranianos, dmi, dmi2}.

The importance of these nanostructures lies in their ability to host robust equilibrium states and topologically protected configurations \cite{Han2009AdvMater31, Streubel2016JPD}, such as skyrmions \cite{Saavedra2025JAC} and flux-closure states, which offer high potential for the development of ultra-low-power data storage technologies (such as racetrack memory) \cite{app1, app3, memoria, Fedorov2024NatComm, memoria2, skirmion1, skirmion2}, magnetic sensors \cite{app2, LACQUANITIAPL2013, seonsor1, seonsor2, seonsor3,dai2024magnet}, neuromorphic computing \cite{bhattacharya2025self} and advanced magnetofluidic devices for targeted drug delivery \cite{biome1, biome2, biome3, biome4, Streubel2016JPD, Three}. In high-symmetry geometries, equilibrium states have been systematically characterized, accounting for curvature and torsion. For example, in cylindrical nanotubes, where curvature is constant and torsion is null, the magnetization adopts axial, azimuthal, or mixed configurations depending on the radius, thickness, and length \cite{Omar}. In these systems, reversal is usually mediated by vortex-type domain walls (VDWs), whose extent is determined by the energetic balance between exchange and dipolar interactions. However, when a transverse symmetry breaking is introduced through eccentricity, double-vortex states \cite{nanodotselip, doblevor} can be stabilized, while in nanohelices where curvature and torsion coexist, quasi-tangential or onion-type states arise depending on the magnitude of curvature and torsion \cite{Yershov2015PRB, Yershov2016PRB, Sheka2015PRB, doblehelice}. 
A recent experimental and micromagnetic study of helicoidal magnetic nanotubes (hereafter referred to as nanoscrews) further showed that the helical geometry can influence the nucleation and propagation of VDW, highlighting the roles of geometric curvature and chirality in the magnetic behavior of three-dimensional nanostructures \cite{screw}. Besides, this study provides a glimpse of the equilibrium states and reversal-mode dynamics, but it is presented only for a constant and large eccentricity.

Although literature considers systems that combine curvature and torsion or curvature and transverse symmetry breaking through eccentricity, the magnetic response of structures that simultaneously combine curvature, torsion, and eccentricity remains little explored. In particular, it has not been established how the combination of curvature, torsion, and eccentricity modifies the balance between dipolar and exchange energy, nor how this competition redefines the stability of remanent states, the reversal mechanisms mediated by VDW, and, consequently, the coercive field. 

\begin{figure}[h]
\includegraphics[width=14.2cm]{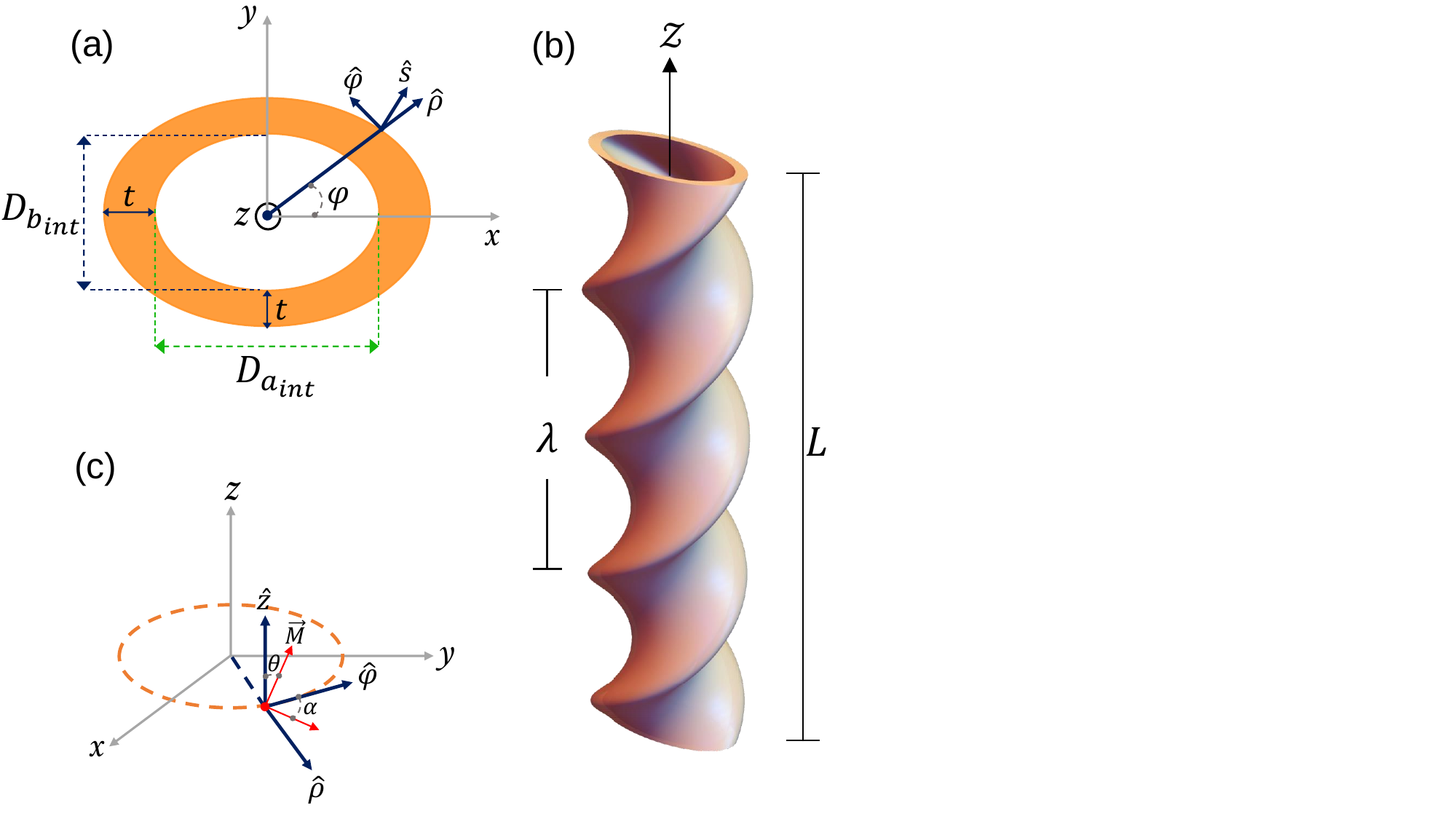}
\caption{\label{tornillo} Illustration of a nanoscrew. (a) Elliptical cross-section of the nanoscrew formed by the larger and smaller inner diameters ($D_{a_\mathrm{int}}$ and $D_{b_{int}}$) with thickness $t$. (b) A nanoscrew of length $L$ and torsion ($\text{w}=L/\lambda$) related to its pitch $\lambda$. (c) System of cylindrical coordinates ($\rho, \varphi, z$) and the magnetization field oriented by the angles $\theta$ and $\alpha$
}
\end{figure}

Therefore, in this work, we explore the equilibrium magnetic states of a scarcely explored geometry in the literature: the magnetic nanoscrew. A system that combines curvature, torsion $(\text{w})$, and eccentricity $(\epsilon)$ in the same magnetic membrane (see Fig.~\ref{tornillo}). Through micromagnetic simulations, we systematically analyze how $\text{w}$, $\epsilon$, the thickness ($t$), and the larger inner diameter ($D_{a_\mathrm{int}}$) govern the hysteresis loops, understanding their role in the coercive field, stability, and transition between equilibrium states at remanence. We show that eccentricity modifies the distribution of surface charges on the elliptical mantle, reducing the characteristic length of the reversal-mode nucleation (as of the VDWs) and increasing the exchange contribution, leading to a systematic increase in the coercive field. In contrast, torsion produces local geometric perturbations during nucleation of reversal modes, keeping the reversal mechanism practically unchanged. Likewise, we identify an energetic degeneracy among the different configurations of the equilibrium mixed state.

The manuscript is organized as follows. In Section II, we describe the micromagnetic simulations and the computational methodology used in this work. Section III presents the results and discussion, including analyses of magnetic equilibrium states, phase diagrams, energetic contributions, and coercive fields as a function of the geometry parameters. Finally, Section IV summarizes the study's main conclusions.

\section{MICROMAGNETIC SIMULATIONS}

In the following, we study the coercive fields and magnetic equilibrium states at remanence of permalloy (NiFe) nanoscrews using the Object-Oriented Micromagnetic Framework (OOMMF) software \cite{DonahueOOMMF}. This micromagnetic solver based on finite differences solves the Landau-Lifshitz-Gilbert (LLG) equation of motion:

\begin{equation}
\dot{\textbf{M}} = -\gamma (\mathbf{M} \times \mathbf{H}_{\text{eff}}) + \alpha \frac{1}{M_s} \left( \mathbf{M} \times \dot{\textbf{M}} \right), \quad
\label{eq:llgEq}
\end{equation}
\noindent where the first term at right hand side (RHS) describes the precessional motion of the magnetization $\mathbf{M}$ around the effective field $\mathbf{H}_{\text{eff}}$. The second term at RHS is the Gilbert damping torque. 

The sample space is discretized into small cubic cells of $2\times 2 \times 5$ nm$^{3}$, with usual permalloy material parameters consisting of a saturation magnetization $M_{s}=796\times 10^{3}$ A m$^{-1}$, exchange stiffness constant $A=13\times 10^{-12}$ J m$^{-1}$ and exchange length $l_{ex}=5.72$ nm.  The simulations were performed for major axis values (larger internal diameters) $D_{a_\mathrm{int}}\in\{40, 60,80\}$ nm; each one with thickness of $10$ nm and $20$ nm, torsion from 0 to 3 in steps of $0.1$, and different eccentricities as shown in Table \ref{tab:table1}.

\begin{table}[h!]
\caption{\label{tab:table1}Values of eccentricity ($\epsilon$) used for each larger internal diameter ($D_{a_\mathrm{int}}$)).
}
\begin{ruledtabular}
\begin{tabular}{lcdr}
\multicolumn{1}{c}{\textrm{$D_{a_\mathrm{int}}$ (nm)}}&
\textrm{$\epsilon$}\\
\colrule
40 &  $0.1$ $0.2$ $0.3$ $0.44$ $0.51$  $0.6$ $0.7$ $0.82$ $0.91$ \\
\hline
60 &  $0.18$ $0.23$ $0.3$ $0.44$ $0.53$ $0.6$ $0.7$ $0.8$ $0.9$ \\
\hline
80 &  $0.16$ $0.2$ $0.3$ $0.41$ $0.51$ $0.6$ $0.7$ $0.81$ $0.91$ \\
\end{tabular}
\end{ruledtabular}
\end{table}

The minor and major axes (smaller and larger inner diameters) $D_{b_\mathrm{int}}$ and $D_{a_\mathrm{int}}$ are related via the eccentricity through the following equation: 

\begin{equation}
D_{b_\mathrm{int}} = D_{a_\mathrm{int}} \sqrt{1 - \epsilon^2}.
\end{equation}

The pitch $\lambda$ is defined as the axial distance along the $\hat z$ direction over which the elliptical cross section of the nanoscrew completes a full $2\pi$ rotation. The total number of such rotations along the NSw length $L$ is quantified by the torsion $\text{w}$. These quantities are related through $\delta=(2\pi\text{w}) z/L$, where  $\delta$  denotes the azimuthal rotation angle of the cross section at position $z$. Accordingly, torsion and pitch are connected by $\text{w}=L/\lambda$.

All simulations were performed for a fixed length $L=4\mu$m, significantly exceeding the transverse dimensions of the elliptical cross section. The ellipticity induces a curvature gradient along the perimeter, which is helicoidally distributed along the NSw axis ($z$ direction), as illustrated in Fig. \ref{tornillo}(a) and (b). The chosen geometrical parameters, including diameter and length, are motivated by values commonly reported for magnetic nanotubes (NTs). In this context, the equilibrium magnetic states of nanotubes \cite{Omar, capProfe} provide a reference framework to assess the impact of eccentricity and torsion on the equilibrium states and coercive fields of NSws.

We simulate the hysteresis loops of the NSw by employing the conjugated gradient method of OOMMF at every applied magnetic field, which implies working at a highly dissipative regime that, in terms of the LLG Eq. \eqref{eq:llgEq}, means not considering the left term at the right-hand side of it. From here, we obtain the equilibrium states at remanence and the coercive fields.
The magnetic field was applied along the NSw $z$ axis and was varied in three intervals: from saturation along the $+\hat z$ direction with 400 mT to 0 mT at steps of $\Delta H = 40$ mT; we continued from 0 mT to $-300$ mT with finer steps of $\Delta H = 6$ mT to observe the magnetization reversal process with higher resolution and determine the coercive field; and finally, we followed from $-300$ mT to $-400$ mT with steps of $\Delta H = 20$ mT to magnetically saturate the system along $-\hat z$  direction. We therefore repeat the procedure in the opposite direction to complete the hysteresis loop. The simulations were carried out in the quasi-static limit, assuming that the applied magnetic field varies slowly enough for the system to reach equilibrium at each field value. To ensure convergence at each step of the cycle, a stopping criterion was imposed, such that the maximum value of $|\mathbf{m} \times \mathbf{H}_{\mathrm{eff}} \times \mathbf{m}|$ throughout the system was lower than $0.01$ A/m, which is a standard procedure in OOMMF simulations. 

For each field value in the hysteresis cycle, the magnetization distribution is saved. After post-processing all this data, the coercive field and equilibrium states at remanence were identified. In particular, the equilibrium states were classified by using the total average of the cylindrical magnetization components $\langle M_{\rho} \rangle$, $\langle M_{\varphi} \rangle$, $\langle M_{z} \rangle$) as order parameters, where $\langle X\rangle$ is the average of $X$ along the NSw volume. This analysis shows that $\langle M_{\rho} \rangle/M_s\ll 1$, $\langle M_{\varphi} \rangle/M_s\ll 1$, and $\langle M_{z} \rangle/M_s\approx 1$, revealing that the remanent magnetic state is almost aligned along the $\hat z$ directions with misalignments nearby the NSw ends. This is consistent with the large length-to-diameter aspect ratio of the NSw and the remanent magnetization states at the limit case of a nanotube. In the next section, we first present an analysis and discussion of the remanent states by examining the textures at the NSw ends, and second, the coercive field. These analyses are done as a function of the NSw eccentricity, torsion, and diameter.

\section{RESULTS AND DICUSSIONS}

To identify the equilibrium states at remanence in the NSw, we analyzed the cylindrical magnetization components averaged per cross-section perpendicular to the $\hat {z} $ axis. Therefore, in the following, we redefine $(\langle M_\rho\rangle,|\langle M_\varphi\rangle|,\langle M_z\rangle)$ as the averages of the magnetization components in cylindrical coordinates at the elliptical NSw cross-section. Notice that we will work with $|\langle M_{\varphi} \rangle|$ instead of $\langle M_{\varphi} \rangle$ to avoid a degeneration in the azimuthal orientation that arises from the conjugated gradient method of OOMMF used to calculate remanent states, as discussed later. In the following, we first analyze and discuss the equilibrium state at remanence and later the coercive field.

\begin{figure*}
\includegraphics[width=18.8cm]{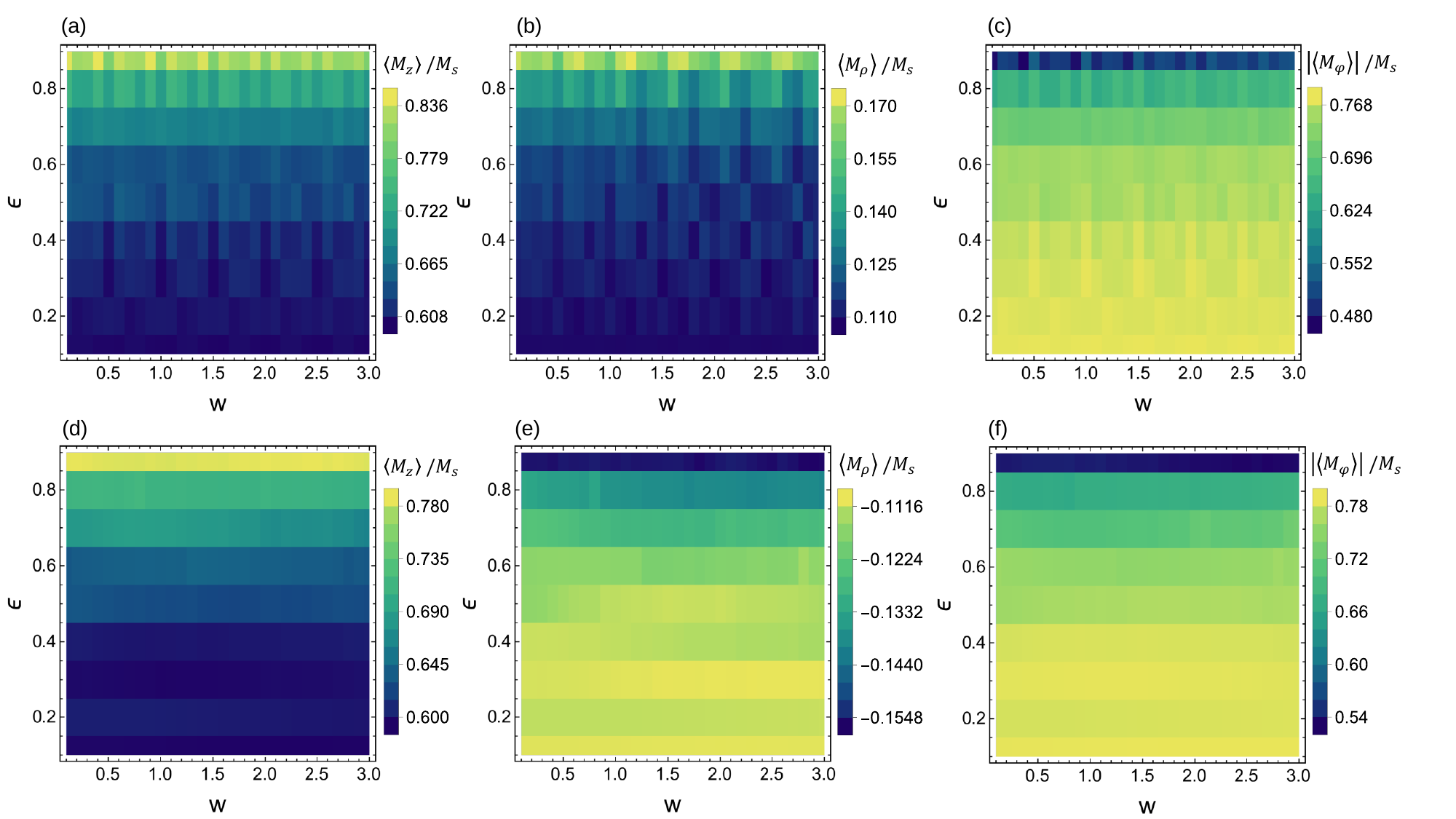} 
\caption{\label{PromST} Cross-section average of the  magnetization field cylindrical components $\langle M_{z} \rangle$, $\langle M_{\rho} \rangle$ and $|\langle M_{\varphi} \rangle|$ at the nanoscrew ends, as function of the eccentricity $\epsilon$ and torsion $\text{w}$ for a nanoscrew with length $L=4 \mu$m, thickness $t=10$ nm and inner diameter $D_{a_\mathrm{int}}=40$ nm. (a)-(c) Averages at the top end. (d)-(f) Average at the bottom end.}
\end{figure*}

\begin{figure*}
\includegraphics[width=18cm]{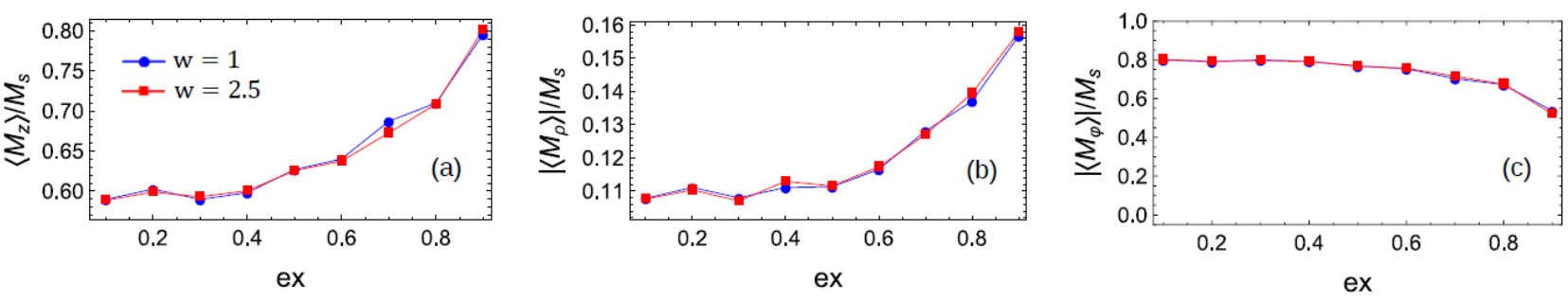} 
\caption{\label{mvsex} Cross-section average of the cylindrical magnetization components $\langle M_{z} \rangle$, $|\langle M_{\rho} \rangle|$ and $|\langle M_{\varphi} \rangle|$ at the nanoscrew ends as a function of the eccentricity for two representative torsion w$=1$ and $2.5$. Nanoscreo with an inner diameter $D_{a_\mathrm{int}}=40$ nm and thickness $t=10$ nm. }
\end{figure*}

In remanence, we found that the magnetization is mostly oriented along the $z$ direction, $\langle M_z\rangle\approx 1$, with non-zero values of $\langle M_{\rho} \rangle$ and $|\langle M_{\varphi} \rangle|$ near the NSw ends. In FIG. \ref{PromST} we show the averages $\langle M_{\rho} \rangle$, $|\langle M_{\varphi} \rangle|$ and $\langle M_{z} \rangle$ at the ends of a NSw with thickness $t=10$ nm and $D_{a_\mathrm{int}}=40$ nm as function of $\epsilon$ and $\text{w}$ (See supplementary material (SM) for larger thickness $t=20$ nm and diameters $D_{a_\mathrm{int}}\in\{60, 80\}$ nm, SM Section I\cite{SupMat}). All our results show remanent states similar to the mixed state in cylindrical nanotubes \cite{Omar,reversion, capProfe}. However, some differences appear related to the gradual (negligible) dependence of the averaged components $(\langle M_{\rho} \rangle,|\langle M_{\varphi} \rangle|,\langle M_{z} \rangle)$ in relation to the eccentricity $\epsilon$  (torsion $\mathrm{w}$). While $|\langle M_{\rho} \rangle|$ remains almost unchanged with eccentricity, the azimuthal  $\langle M_{\varphi} \rangle$ (longitudinal $\langle M_{z} \rangle$) component decreases (increases) as the eccentricity increases. It is more clearly shown in FIG. \ref{mvsex} at two different torsions. Increasing the eccentricity also increases the surface magnetostatic charges at the mantle near the NSw ends due to the azimuthal magnetization component. These charges, defined as $\sigma_\varphi=M_\varphi\ \hat \varphi\cdot\hat s$, with $\hat s$ the unit vector perpendicular to the elliptical mantle, are located mainly in regions of the elliptical mantle with greater curvature. Consequently, increasing the eccentricity increases the ellipticity, the magnetic charges, and their associated demagnetizing field, thereby increasing the dipolar self-interaction. To mitigate the increase of the demagnetizing field with eccentricity, the system reduces the azimuthal component \( |\langle M_\varphi \rangle| \) while enhancing the longitudinal and radial components, \( \langle M_z \rangle \) and \( \langle M_\rho \rangle \), respectively. However, the growth of \( \langle M_\rho \rangle \) is constrained to relatively small values in order to limit the associated surface magnetic charges at the NSw mantle (\( \sigma_s = M_\rho\, \hat{\rho} \cdot \hat{s} \)) and, consequently, its self-demagnetizing field.

\begin{figure*}
\includegraphics[width=18cm]{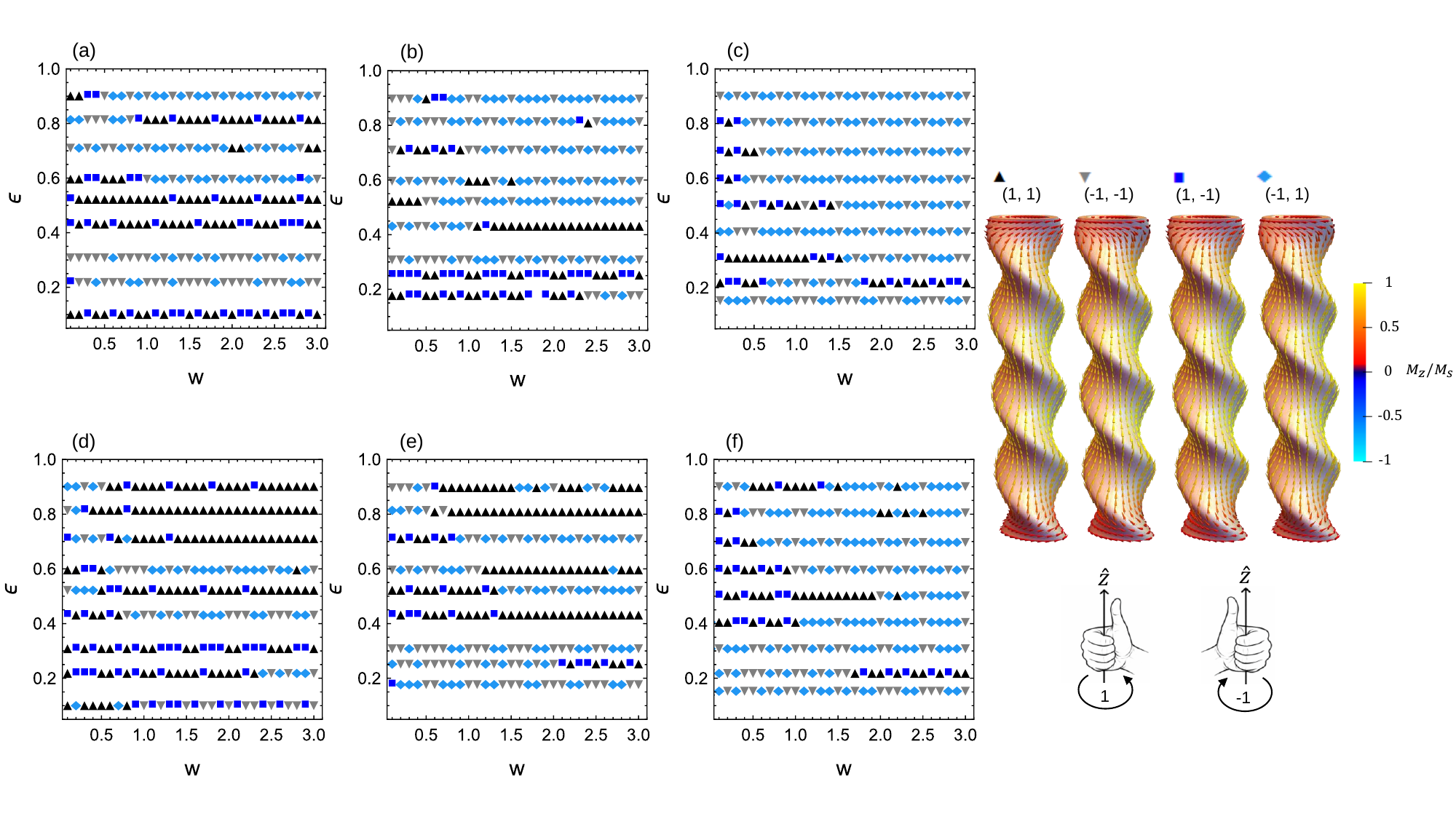}
\caption{Phase diagram of the equilibrium magnetization states of the nanoscrew in terms of its torsion (w) and eccentricity ($\epsilon$). A nanoscrew of length $L=4$ $\mu$m and two thicknesses: (a)-(c) $t=10$nm and (d)-(e) $t=20$nm; and diameters: (a) and (d) $D_{a_\mathrm{int}}=40$nm,  (b) and (e) $D_{a_\mathrm{int}}=60$nm, (c) and (f) $D_{a_\mathrm{int}}=80$nm. Illustration of the  mixed magnetization state with the four configurations according the magnetic vorticity at the nanoscrew ends: black triangle (1,1); gray inverted triangle (-1,-1)); blue full square (1,-1); and blue empty square (-1,1) follow the notation $(a,b)$, where $a$($b$) denotes the magnetization vorticity at the top (bottom) NSw end with $a=\pm 1$ ($b=\pm 1$), where right(left) handed vorticity is denoted with +1(-1) as illustrated.}
\label{diagrama}
\end{figure*}

\begin{figure}[h]
\includegraphics[width=8.5cm]{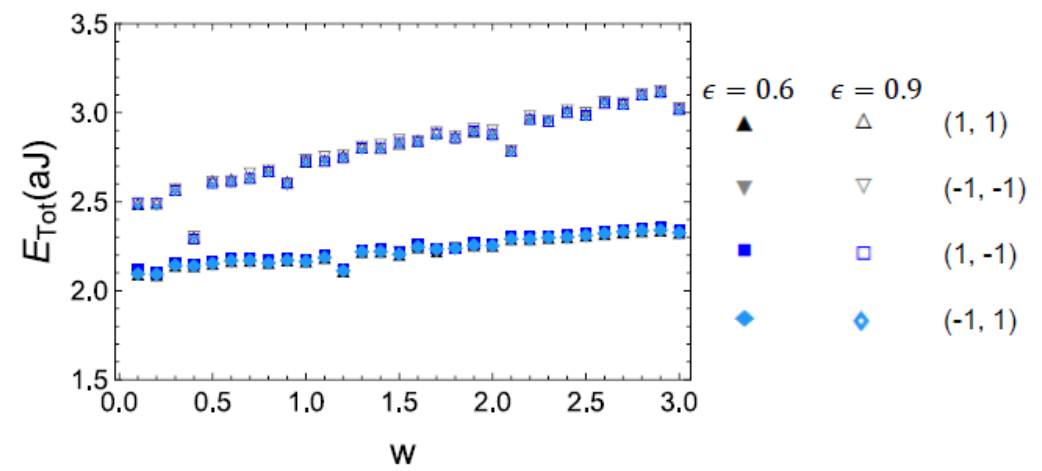} 
\caption{\label{energia} Total magnetic energy of the mixed magnetic state in its four configurations at two eccentricities $\epsilon\in\{0.6,0.91\}$, and as a function of the torsion $\mathrm{w}$. A nanoscrew with $L=4$ $\mu$m, $D_{a_\mathrm{int}}=40$nm and thickness $t=10$nm.
}
\end{figure}

The mixed magnetization state in the NSw can be found in four different configurations depending on the vorticity at both ends of the system. Vorticity is a well-defined physical concept for quantifying fluid rotation and its direction \ cite {ohkitani2010elementary}. However, in this work, we use it intuitively via a handedness rule to understand the magnetization rotation direction at the NSw ends: with the thumb pointing along the $z$ axis, the vorticity is defined as the direction of rotation of the remaining four fingers when they curl to form a fist. In this sense, the left and right hands have opposite vorticities. Denoting the right (left) hand vorticity by +1(-1) number, we use the notation $(a,b)$ to specify the magnetization vorticity at the NSw ends, where $a$ takes only two values with $a=+1$ ($a=-1$) to denote the vorticity at the top end with $M_\varphi\parallel\hat\varphi$ ($M_\varphi\parallel-\hat\varphi$). The same applies to $b$ to describe the vorticity at the bottom end. It gives four possible equilibrium states depending on the magnetization vorticity at the NSw ends, as illustrated in the rightmost pictures in FIG. \ref{diagrama}. In FIG. In \ref{diagrama}(a-f), we show the phase diagrams of the magnetic mixed-state configuration as a function of eccentricity and torsion. We found that the four configurations of the mixed magnetization state do not exhibit a regular pattern in their vorticity as a function of the geometrical parameters, indicating that the four configurations are degenerate in energy. To verify this degeneracy, we calculated the magnetic energy by performing simulations of NSw with predefined states close to the mixed equilibrium states in all four configurations and relaxed the system at a set of two eccentricities $\epsilon\in\{0.6,0.91\}$ and torsions running from 0.1 to 3.0 in steps of 0.1.  The predefined states consisted of an NSw with magnetization aligned to the $z$ axis, whereas the magnetization lay down with $M_z=M_\rho=0$ and $M_\varphi=\pm M_s$  in a region of 10 nm length from the NSw ends. By setting the vorticity by hand for these predefined states and then relaxing, the magnetization reaches the mixed state, keeping its predefined vorticity. FIG. \ref{energia} shows the results of the simulated total magnetic energy (exchange and dipolar energies) of the relaxed states, where one can see that the four mixed state configurations are degenerate at a given eccentricity $\epsilon$ and torsion $\mathrm{w}$, in an NSw with $D_{a_\mathrm{int}}=40$ nm.

\begin{figure*}
\includegraphics[width=18cm]{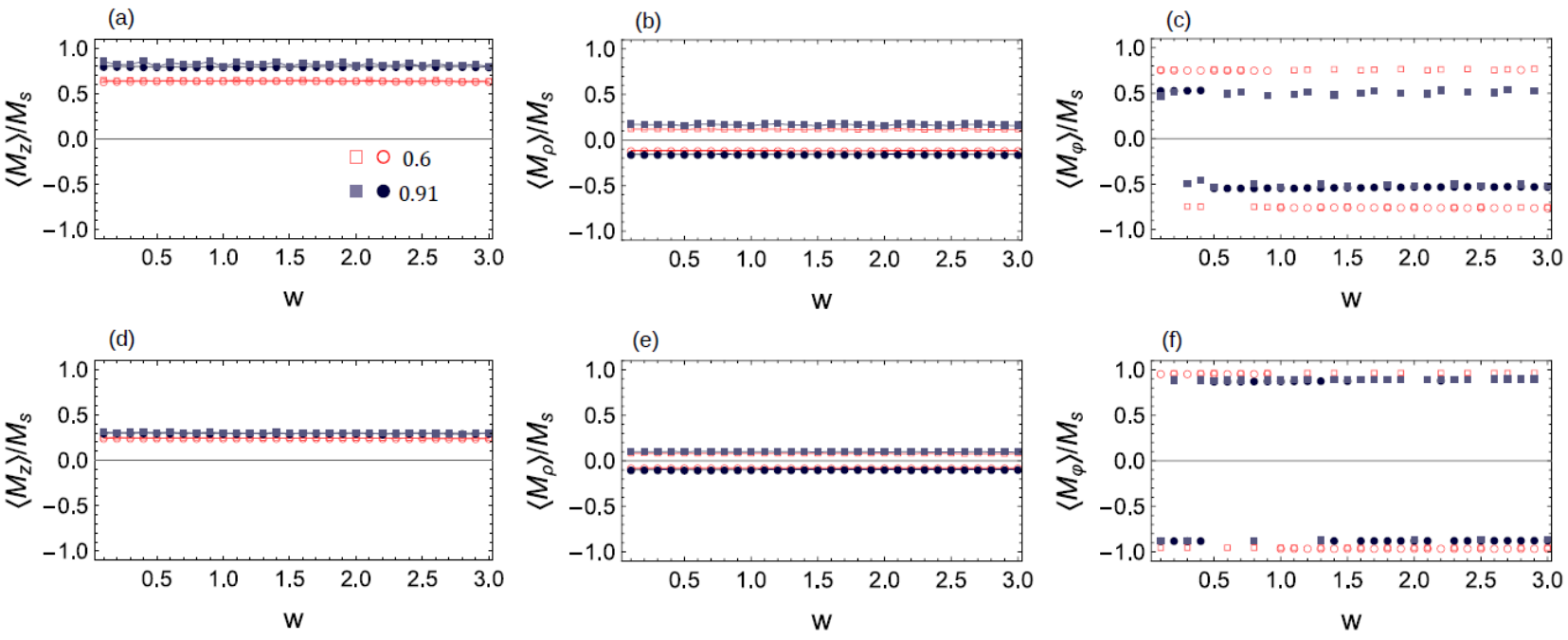} 
\caption{\label{Estado06y09} Cross-section average of the magnetization cylindrical components $\langle M_{z} \rangle$, $\langle M_{\rho} \rangle$ and $\langle M_{\varphi} \rangle$ at the nanoscrew ends as  function of the torsion $\text{w}$ and two eccentricities $\epsilon\in\{0.6,0.91\}$. (a), (b) and (c) are for an inner diameter $D_{a_\mathrm{int}}=40$ nm and thickness $t=10$ nm. (d), (e) and (f) are an inner diameter $D_{a_\mathrm{int}}=80$ nm and thickness $t=20$ nm. The square (round) symbol denotes the averages at the top (bottom) end. }
\end{figure*}

In FIG. \ref{Estado06y09}, we show the cross-section averaged cylindrical magnetization components at both ends for two representative eccentricities $\epsilon\in\{0.6, 0.91\}$ as a function of the simulated torsion values. Three features are worth of being noticed: first, as the eccentricity increases, the axial magnetization $\langle M_z\rangle$ increases and the azimuthal magnetization $\langle M_\varphi\rangle$ decreases, which, as explained previously, is an effect emerging from modifications in dipolar interaction as the ellipciticy of the NSw increases with the eccentricity; second, $M_{\varphi}$ increases as the diameter also increases (see FIG. \ref{Estado06y09}(d)-(f)), which is consequent with a strengthening of the dipolar interaction due to the diameter increase. As a response, the magnetization acquires an additional tendency to lie along the azimuthal direction to increase the flux-closure condition, thus tending to minimize the dipolar interaction.  Finally, as a third, the four configurations of the mixed magnetization state do not show a regular order in their vorticity as a function of the geometrical parameters $\epsilon$ and $\text{w}$, which is an indication that the four configurations are degenerated in energy.

Another key magnetic property of the NSw is the coercive field as a function of its geometric parameters. The coercive field corresponds to the magnitude of the applied magnetic field required to reduce the magnetization from saturation to zero, i.e., the field at which the magnetization vanishes in the hysteresis loop. In our case, when the magnetic field is applied along the $\hat {z} $ axis, the NSw exhibits an almost rectangular hysteresis loop, as shown in SM Section II \cite{SupMat}. Such rectangular loops are characteristic of bistable magnetic systems, where the magnetization evolves quasi-statically between two nearly unchanged remanent states. For the NSw, the magnetization switches between two mixed states with axial components aligned parallel ($M_z\parallel \hat z$) and antiparallel ($M_z\parallel -\hat z$) to the axis. Our simulations indicate that this transition is governed by a vortex-domain-wall (VDW) reversal mechanism. In particular, the nucleation of the VDW at the onset of reversal plays a decisive role in determining the coercive field $H_c$. Notably, the identification of VDW-mediated reversal is consistent with recent experimental observations in nanoscrews \cite{screw}, as well as with the limiting case of magnetic nanotubes ($\epsilon\approx0$) having the same length, diameter, and thickness as the NSw \cite{Omar,reversion, capProfe}.
In the following, we analyze the influence of eccentricity and torsion on the fundamental characteristics of the VDW at the onset of reversal to elucidate the role of NSw geometry in determining the coercive field.

Within the range of geometrical parameters considered here, the magnetization reversal mechanism in the NSw is found to be identical to that of a nanotube. As the applied field is reduced from saturation along the $\hat {z} $ axis to values approaching the coercive field, vortex-domain walls (VDWs) nucleate at both ends of the NSw. When the applied field reaches the coercive value, the VDWs acquire sufficient energy to detach from the ends and propagate along the length of the structure. Eventually, the two VDWs meet within the NSw and annihilate, resulting in complete magnetization reversal. Accordingly, the coercive field is determined by the applied field at which the VDWs become mobile and propagate through the system. Therefore, understanding the influence of the NSw geometrical parameters on VDW nucleation is essential to elucidate the roles of eccentricity, torsion, and structural dimensions in setting the coercive field.

\begin{figure*}[th!]
\includegraphics[width=18cm]{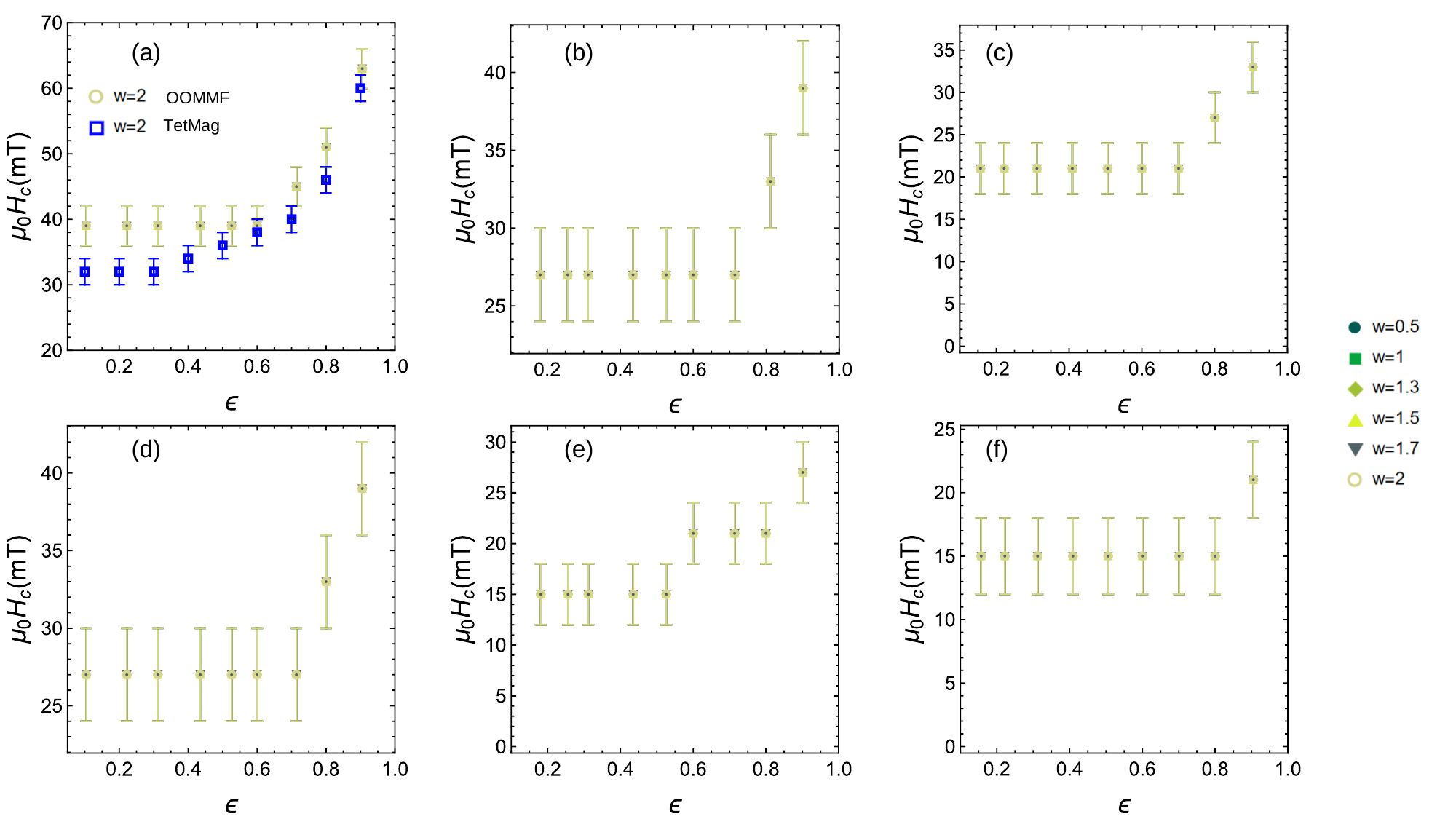} 
\caption{\label{hc} Coercive field $H_{c}$ as function of the eccentricity $\epsilon$ for a nanoscrew with length $L=4$ $\mu$m, and diameter (a),(d) $D_{a_\mathrm{int}}=40$nm, (b),(e) $D_{a_\mathrm{int}}=60$nm and (c),(e) $D_{a_\mathrm{int}}=80$nm.  Coercive field for thickness (a)-(c) $t=10$nm and (d)-(e) $t=20$ nm.}
\end{figure*}

In Fig. \ref{hc}, we summarize the dependence of the coercive field on the NSw eccentricity, torsion, inner diameter, and two representative thicknesses. Although most results were obtained using OOMMF simulations, we benchmark the coercive field against results computed with \textit{TetMag}\cite{tetmagHertel}, a micromagnetic finite-element solver, for the case of a NSw with \( D_{a_{\text{int}}} = 40 \), as shown in Fig.~\ref{hc}(a). Both approaches exhibit the same overall trend, with \textit{TetMag} systematically yielding lower coercive fields than OOMMF. This discrepancy can be attributed to the discretization scheme: the finite-difference mesh in OOMMF introduces an intrinsic roughness at the NSw mantle, whereas the finite-element mesh in \textit{TetMag} provides a smoother surface representation. In particular, the roughness at the NSw mantle?especially near the VDW nucleation region?acts as a pinning site for the domain wall. As a result, larger applied fields are required to depin the VDW from the NSw ends and trigger reversal, leading to higher coercive fields in OOMMF simulations. Despite these quantitative differences, the good qualitative agreement between \textit{TetMag} and OOMMF supports the validity of our results. The error bars (6 mT) reflect the field-step resolution used in the simulations. The coercive field is found to increase with eccentricity, while remaining essentially insensitive to torsion. Notably, this increase becomes pronounced for eccentricities larger than $0.6$, corresponding to structures in which the larger inner diameter exceeds the smaller inner diameter by approximately 30\%. This behavior can be understood by considering the effects of eccentricity on the vortex-domain wall (VDW) at the onset of reversal. The VDW constitutes the preferred reversal mode as it promotes flux closure and minimizes dipolar energy. However, during its formation, surface magnetic charges ($\sigma =\mathbf{M}_\varphi\cdot \hat s$) develop on the elliptical mantle, generating a demagnetizing field that opposes the VDW nucleation. To mitigate this effect, the VDW contracts, adopting a length smaller than that typically observed in nanotubes ($\sim 50$ nm), thereby reducing the associated surface charges. This contraction enhances the exchange energy, thereby increasing the coercive field. The effect becomes more pronounced with increasing eccentricity, as larger eccentricities give rise to stronger dipolar charge accumulation near the ends and, consequently, larger demagnetizing fields. As a result, the VDW undergoes a more significant contraction, ultimately yielding higher coercive fields.

A similar argument applies to the role of torsion in determining the coercive field. As shown in Fig. \ref{hc}, the VDW remains essentially unchanged even at the largest torsion considered. For a NSw of length 4 $\mu$m and maximum torsion $\text{w}=3$, the azimuthal rotation of the structure within the region relevant for VDW nucleation ($\sim 50$ nm) is only $\Delta \delta \approx 6.7$ degrees. Since the VDW length at the onset of reversal is smaller than this characteristic length scale, the structural variation experienced by the VDW is minimal and can be regarded as a weak perturbation. Consequently, both the nucleation and subsequent propagation of the VDW remain largely unaffected by torsion within the studied range, leading to a negligible impact on the coercive field.

In addition, FIG.\ref{hc} also shows a reduction on the coercive field by increasing the NSw diameter and thickness. For the smallest diameter, such as $D_{a_\mathrm{int}}=40$nm (see Fig. \ref{hc} (a)-(c)), the exchange interaction acquire more relevance in relation to the dipolar interaction, thus leading to larger coercive fields than for largest diameters such as $D_{a_\mathrm{int}}=80$ nm where the dipolar has gained influence. 

\section{CONCLUSIONS}

In this work, we have investigated the equilibrium magnetization states and reversal processes in nanoscrew geometries that combine curvature, torsion, and cross-sectional eccentricity within a single magnetic membrane. Our results demonstrate that the remanent configuration is governed by the competition between exchange and dipolar interactions, leading to a mixed state across the full range of geometrical parameters considered, including thickness, diameter, eccentricity, and torsion. This mixed state supports four degenerate vorticity configurations, indicating that the combined presence of torsion and eccentricity does not favor any particular global vorticity.

Magnetization reversal proceeds via vortex-domain-wall (VDW) propagation, consistent with the cylindrical nanotube limit. However, eccentricity has a pronounced impact on the VDW structure. Increasing eccentricity enhances surface magnetostatic charges on the elliptical mantle, which induces a reduction of the VDW axial extent at nucleation. This contraction increases the exchange energy contribution, leading to a systematic increase in the coercive field. In contrast, torsion introduces only weak geometric perturbations at the characteristic VDW length scale, resulting in negligible changes to the reversal mechanism and a minimal dependence of the coercive field on $\mathrm{w}$.

\section*{Declaration of competing interest}
The authors declare that they have no known competing financial interests or personal relationships that could have appeared to influence the work reported in this paper.

\begin{acknowledgments}
  This research has received funding support from Chilean Doctorado Nacional ANID via fellowship Grant 21221865.
\end{acknowledgments}

\section*{Data Availability Statement}

The data are available upon reasonable request from the authors.

\section*{CRediT authorship contribution statement}
 \textbf{Victoria Acosta-Pareja}: Writing - original draft, Visualization, Conceptualization, Investigation, Formal Analysis.  \textbf{Valeria M. A. Salinas}: Writing - original draft, Visualization, Formal Analysis. \textbf{Omar J. Suarez}: Writing - review . \textbf{Attila K\'akay}: Writing - review \& editing, Conceptualization, Investigation. \textbf{Jorge A. Ot\'alora}: Writing - review \& editing, Conceptualization, Supervision, Resources, Project administration, Methodology, Investigation, Formal Analysis.

\appendix

\nocite{*}

\bibliography{References}

@PREAMBLE{
 "\providecommand{\noopsort}[1]{}" 
 # "\providecommand{\singleletter}[1]{#1}%" 
}

@article{app1,
  title={Controlling domain wall thermal stability switching in magnetic nanowires for storage memory nanodevices},
  author={Al Bahri, M},
  journal={Journal of Magnetism and Magnetic Materials},
  volume={543},
  pages={168611},
  year={2022},
  publisher={Elsevier}
}

@article{app2,
  title={Monolithic sensor integration in CMOS technologies},
  author={Fernandez, Daniel and Michalik, Piotr and Valle, Juan and Banerji, Saoni and Sanchez-Chiva, Josep Maria and Madrenas, Jordi},
  journal={IEEE Sensors Journal},
  volume={23},
  number={2},
  pages={1479--1496},
  year={2022},
  publisher={IEEE}
}

@article{app3,
  title={Ultrahigh density vertical magnetoresistive random access memory},
  author={Zhu, Jian-Gang and Zheng, Youfeng and Prinz, Gary A},
  journal={Journal of Applied Physics},
  volume={87},
  number={9},
  pages={6668--6673},
  year={2000},
  publisher={American Institute of Physics}
}

@article{memoria,
  title={Magnetic domain-wall racetrack memory},
  author={Parkin, Stuart SP and Hayashi, Masamitsu and Thomas, Luc},
  journal={science},
  volume={320},
  number={5873},
  pages={190--194},
  year={2008},
  publisher={American Association for the Advancement of Science}
}

@article{Fedorov2024NatComm,
  title={Self-assembly of Co/Pt stripes with currentinduced domain wall motion towards 3D racetrack devices},
  author={Fedorov, Pavel and Soldatov, Ivan and Neu, Volker and Sch{\"a}fer, Rudolf and Schmidt, Oliver G and Karnaushenko, Daniil},
  journal={nature communications},
  volume={15},
  number={1},
  pages={2048},
  year={2024},
  publisher={Nature Publishing Group UK London}
}

@article{memoria2,
  title={Three-dimensional racetrack memory devices designed from freestanding magnetic heterostructures},
  author={Gu, Ke and Guan, Yicheng and Hazra, Binoy Krishna and Deniz, Hakan and Migliorini, Andrea and Zhang, Wenjie and Parkin, Stuart SP},
  journal={Nature nanotechnology},
  volume={17},
  number={10},
  pages={1065--1071},
  year={2022},
  publisher={Nature Publishing Group UK London}
}

@article{seonsor1,
  title={Sensitivity and noise of a magnetic field sensor based on magnetostatic spin wave YIG device and its integrated electronics},
  author={Haas, Olivier and Dufay, Basile and Saez, S{\'e}bastien and Dolabdjian, Christophe},
  journal={IEEE Sensors Journal},
  volume={20},
  number={23},
  pages={14148--14156},
  year={2020},
  publisher={IEEE}
}

@article{seonsor2,
  title={Four-port Characterization of YIG Magnonic Device: a Way to Improve Magnetic Sensors based on YIG Device},
  author={Da, BFE and Dufay, B and Saez, S},
  journal={IEEE Transactions on Magnetics},
  year={2025},
  publisher={IEEE}
}

@article{seonsor3,
  title={Development of a magnonic-based magnetic sensor: Comparison of two different implementations with YIG material},
  author={Haas, O and Dufay, B and Saez, S},
  journal={IEEE Transactions on Magnetics},
  volume={59},
  number={2},
  pages={1--6},
  year={2023},
  publisher={IEEE}
}

@article{biome1,
  title={Hyperthermia in low aspect-ratio magnetic nanotubes for biomedical applications},
  author={Gutierrez-Guzman, DF and Lizardi, LI and Ot{\'a}lora, JA and Landeros, P},
  journal={Applied Physics Letters},
  volume={110},
  number={13},
  year={2017},
  publisher={AIP Publishing}
}

@article{biome2,
  title={Synthesis of hybrid magneto-plasmonic nanoparticles with potential use in photoacoustic detection of circulating tumor cells},
  author={Ovejero, Jesus G and Yoon, Soon Joon and Li, Junwei and Mayoral, Alvaro and Gao, Xiaohu and O’Donnell, Matthew and Garc{\'\i}a, Miguel A and Herrasti, Pilar and Hernando, Antonio},
  journal={Microchimica Acta},
  volume={185},
  number={2},
  pages={130},
  year={2018},
  publisher={Springer}
}

@article{biome3,
  title={Gold nanoparticles immobilized on halloysite nanotubes for spatially-temporally localized photohyperthermia},
  author={Kornilova, AV and Kuralbayeva, GA and Stavitskaya, AV and Gorbachevskii, MV and Karpukhina, OV and Lysenko, IV and Pryadun, VV and Novikov, AA and Vasiliev, AN and Timoshenko, V Yu},
  journal={Applied Surface Science},
  volume={566},
  pages={150671},
  year={2021},
  publisher={Elsevier}
}

@article{biome4,
  title={Potential applications of boron nitride nanotubes as drug delivery systems},
  author={Ciofani, Gianni},
  journal={Expert opinion on drug delivery},
  volume={7},
  number={8},
  pages={889--893},
  year={2010},
  publisher={Taylor \& Francis}
}

@article{Three,
  title={Three-dimensional nanomagnetism},
  author={Fern{\'a}ndez-Pacheco, Amalio and Streubel, Robert and Fruchart, Olivier and Hertel, Riccardo and Fischer, Peter and Cowburn, Russell P},
  journal={Nature communications},
  volume={8},
  number={1},
  pages={15756},
  year={2017},
  publisher={Nature Publishing Group UK London}
}

@article{dmi,
  title={Dzyaloshinskii-Moriya domain walls in magnetic nanotubes},
  author={Goussev, Arseni and Robbins, Jonathan M and Slastikov, Valeriy and Tretiakov, Oleg A},
  journal={Physical Review B},
  volume={93},
  number={5},
  pages={054418},
  year={2016},
  publisher={APS}
}

@article{dmi2,
  title={Mesoscale Dzyaloshinskii-Moriya interaction: geometrical tailoring of the magnetochirality},
  author={Volkov, Oleksii M and Sheka, Denis D and Gaididei, Yuri and Kravchuk, Volodymyr P and R{\"o}{\ss}ler, Ulrich K and Fassbender, J{\"u}rgen and Makarov, Denys},
  journal={Scientific reports},
  volume={8},
  number={1},
  pages={866},
  year={2018},
  publisher={Nature Publishing Group UK London}
}

@article{vale,
  title={Azimuthal anisotropy induced by partial flux-closure in self-assembled tubular permalloy membranes},
  author={Singh, Balram and Salinas, Valeria MA and Loeffler, Markus and Soldatov, Ivan and Rivkin, Boris and Hantusch, Martin and Rellinghaus, Bernd and Sch{\"a}fer, Rudolf and Ot{\'a}lora, Jorge A and Neu, Volker},
  journal={npj Flexible Electronics},
  volume={9},
  number={1},
  pages={89},
  year={2025},
  publisher={Nature Publishing Group UK London}
}

@article{chesun,
  title={Magnetic properties of a long, thin-walled ferromagnetic nanotube},
  author={Sun, Chen and Pokrovsky, Valery L},
  journal={Journal of magnetism and magnetic materials},
  volume={355},
  pages={121--130},
  year={2014},
  publisher={Elsevier}
}

@article{Mobius,
  title={Coupling of chiralities in spin and physical spaces: The M{\"o}bius ring as a case study},
  author={Pylypovskyi, Oleksandr V and Kravchuk, Volodymyr P and Sheka, Denis D and Makarov, Denys and Schmidt, Oliver G and Gaididei, Yuri},
  journal={Physical Review Letters},
  volume={114},
  number={19},
  pages={197204},
  year={2015},
  publisher={APS}
}

@article{Omar,
  title={Equilibrium states and vortex domain wall nucleation in ferromagnetic nanotubes},
  author={Landeros, P and Suarez, OJ and Cuchillo, A and Vargas, P},
  journal={Physical Review B—Condensed Matter and Materials Physics},
  volume={79},
  number={2},
  pages={024404},
  year={2009},
  publisher={APS}
}

@article{Yershov2015PRB,
   author       = "K. V. Yershov and V. P. Kravchuk and D. D. Sheka and Y. Gaididei",
   title        = "Curvature induced domain wall pinning in a magnetic nanowire",
   year         = "2015",
   journal      = "Phys.\ Rev.\ B",
   volume       = "92",
   pages        = "104412",
}

@article{Yershov2016PRB,
   author       = "K. V. Yershov and V. P. Kravchuk and D. D. Sheka and Y. Gaididei",
   title        = "Curvature and torsion effects in spin-current driven domain wall motion in a helix magnetic nanowire",
   year         = "2016",
   journal      = "Phys.\ Rev.\ B",
   volume       = "93",
   pages        = "094418",
}

@article{Sheka2015PRB,
   author       = "D. D. Sheka and V. P. Kravchuk and K. V. Yershov and Y. Gaididei",
   title        = "Torsion-induced effects in magnetic nanowires with easy-tangential anisotropy",
   year         = "2015",
   journal      = "Phys.\ Rev.\ B",
   volume       = "92",
   pages        = "054417",
}

@article{doblehelice,
  title={Complex free-space magnetic field textures induced by three-dimensional magnetic nanostructures},
  author={Donnelly, Claire and Hierro-Rodr{\'\i}guez, Aurelio and Abert, Claas and Witte, Katharina and Skoric, Luka and Sanz-Hern{\'a}ndez, D{\'e}dalo and Finizio, Simone and Meng, Fanfan and McVitie, Stephen and Raabe, J{\"o}rg and others},
  journal={Nature nanotechnology},
  volume={17},
  number={2},
  pages={136--142},
  year={2022},
  publisher={Nature Publishing Group UK London}
}

@article{screw,
  title={Curvilinear magnetic effects in helicoid nanotubes},
  author={Fullerton, John and Phatak, Charudatta},
  journal={npj Spintronics},
  volume={4},
  number={1},
  pages={10},
  year={2026},
  publisher={Nature Publishing Group UK London}
}

@article{reversion,
  title={Reversal modes in magnetic nanotubes},
  author={Landeros, P and Allende, S and Escrig, J and Salcedo, E and Altbir, D and Vogel, EE},
  journal={Applied Physics Letters},
  volume={90},
  number={10},
  year={2007},
  publisher={AIP Publishing}
}

@article{Han2009AdvMater31,
  title={Structural and magnetic properties of various ferromagnetic nanotubes},
  author={Han, Xiu-Feng and Shamaila, Shahzadi and Sharif, Rehana and Chen, Jun-Yang and Liu, Hai-Rui and Liu, Dong-Ping},
  journal={Advanced Materials},
  volume={21},
  number={45},
  pages={4619--4624},
  year={2009},
  publisher={Wiley Online Library}
}

@article{ucranianos,
  title={Curvature effects on phase transitions in chiral magnets},
  author={Yershov, Kostiantyn and Kravchuk, Volodymyr and Sheka, Denis and Roessler, Ulrich},
  journal={SciPost Physics},
  volume={9},
  number={4},
  pages={043},
  year={2020}
}

@article{skirmion1,
  title={A strategy for the design of skyrmion racetrack memories},
  author={Tomasello, Riccardo and Martinez, E and Zivieri, Roberto and Torres, Luis and Carpentieri, Mario and Finocchio, Giovanni},
  journal={Scientific reports},
  volume={4},
  number={1},
  pages={6784},
  year={2014},
  publisher={Nature Publishing Group UK London}
}

@article{skirmion2,
  title={Reversible writing/deleting of magnetic skyrmions through hydrogen adsorption/desorption},
  author={Chen, Gong and Ophus, Colin and Quintana, Alberto and Kwon, Heeyoung and Won, Changyeon and Ding, Haifeng and Wu, Yizheng and Schmid, Andreas K and Liu, Kai},
  journal={Nature communications},
  volume={13},
  number={1},
  pages={1350},
  year={2022},
  publisher={Nature Publishing Group UK London}
}

@article{Saavedra2025JAC,
  title={Impact of geometry on magnetic textures in FeGe nanowires: Fractional skyrmion-vortex states, twisted skyrmions, and helical states},
  author={Saavedra, Eduardo and Pereira, Alejandro and D{\'\i}az, Pablo and Escrig, Juan and Bajales, Noelia and Valdez, Lucy A},
  journal={Journal of Alloys and Compounds},
  pages={182703},
  year={2025},
  publisher={Elsevier}
}

@article{nanodotselip,
  title={Phase diagram of magnetic configurations for soft magnetic nanodots of circular and elliptical shape obtained by micromagnetic simulation},
  author={Novais, ERP and Guimaraes, AP},
  journal={arXiv preprint arXiv:0909.5686},
  year={2009}
}

@article{doblevor,
  title={Scaling analysis and application: phase diagram of magnetic nanorings and elliptical nanoparticles},
  author={Zhang, Wen and Singh, Rohit and Bray-Ali, Noah and Haas, Stephan},
  journal={Physical Review B—Condensed Matter and Materials Physics},
  volume={77},
  number={14},
  pages={144428},
  year={2008},
  publisher={APS}
}

@article{Streubel2016JPD,
  title={Magnetism in curved geometries},
  author={Streubel, Robert and Fischer, Peter and Kronast, Florian and Kravchuk, Volodymyr P and Sheka, Denis D and Gaididei, Yuri and Schmidt, Oliver G and Makarov, Denys},
  journal={Journal of Physics D: Applied Physics},
  volume={49},
  number={36},
  pages={363001},
  year={2016},
  publisher={IOP Publishing}
}

@BOOKLET{DonahueOOMMF,
   author = "M. J. Donahue and D. Porter",
   title = "OOMMF User's  Guide, Version 1.0",
   howpublished = "National Institute of Standards and Technology",
   year = "1999",
}

@incollection{capProfe,
  title={Tubular geometries},
  author={Landeros, Pedro and Ot{\'a}lora, Jorge A and Streubel, Robert and K{\'a}kay, Attila},
  booktitle={Curvilinear Micromagnetism: From Fundamentals to Applications},
  pages={163--213},
  year={2022},
  publisher={Springer}
}

@book{ohkitani2010elementary,
  title={An Elementary Account of Vorticity and Related Equations},
  author={Ohkitani, K},
  year={2010},
  publisher={Cambridge University Press}
}

@article{LACQUANITIAPL2013,
  title={Three-dimensional spin nanosensor based on reliable tunnel Josephson nano-junctions for nanomagnetism investigations},
  author={Granata, Carmine and Vettoliere, Antonio and Russo, Roberto and Fretto, Matteo and De Leo, Natascia and Lacquaniti, Vincenzo},
  journal={Applied Physics Letters},
  volume={103},
  number={10},
  year={2013},
  publisher={AIP Publishing}
}

@article{li2023magnetic,
  title={Magnetic characterization techniques and micromagnetic simulations of magnetic nanostructures: from zero to three dimensions},
  author={Li, Xin and Wang, Zhaolian and Lei, Zhongyun and Ding, Wei and Shi, Xiao and Yan, Jujian and Ku, Jiangang},
  journal={Nanoscale},
  volume={15},
  number={48},
  pages={19448--19468},
  year={2023},
  publisher={Royal Society of Chemistry}
}

@article{APLMATPACHECO2020,
  title={Launching a new dimension with 3D magnetic nanostructures},
  author={Fischer, Peter and Sanz-Hern{\'a}ndez, D{\'e}dalo and Streubel, Robert and Fern{\'a}ndez-Pacheco, Amalio},
  journal={APL Materials},
  volume={8},
  number={1},
  year={2020},
  publisher={AIP Publishing}
}

@article{chumak2022advances,
  title={Advances in magnetics roadmap on spin-wave computing},
  author={Chumak, Andrii V and Kabos, Pavel and Wu, Mingzhong and Abert, Claas and Adelmann, Christoph and Adeyeye, Adekunle Olusola and {\AA}kerman, Johan and Aliev, Farkhad G and Anane, Abdelmadjid and Awad, Ahmad and others},
  journal={IEEE Transactions on Magnetics},
  volume={58},
  number={6},
  pages={1--72},
  year={2022},
  publisher={IEEE}
}

@article{Gubbiotti_2025,
  title={2025 roadmap on 3D nanomagnetism},
  author={Gubbiotti, Gianluca and Barman, Anjan and Ladak, Sam and Bran, Cristina and Grundler, Dirk and Huth, Michael and Plank, Harald and Schmidt, Georg and Van Dijken, Sebastiaan and Streubel, Robert and others},
  journal={Journal of Physics: Condensed Matter},
  volume={37},
  number={14},
  pages={143502},
  year={2025},
  publisher={IOP Publishing}
}

@article{bhattacharya2025self,
  title={Self-assembled 3D Interconnected Magnetic Nanowire Networks for Neuromorphic Computing},
  author={Bhattacharya, Dhritiman and Langton, Colin and Rajib, Md Mahadi and Marlowe, Erin and Chen, Zhijie and Al Misba, Walid and Atulasimha, Jayasimha and Zhang, Xixiang and Yin, Gen and Liu, Kai},
  journal={ACS applied materials \& interfaces},
  volume={17},
  number={13},
  pages={20087--20095},
  year={2025},
  publisher={ACS Publications}
}

@article{dai2024magnet,
  title={A Magnet Splicing Method for Constructing a Three-Dimensional Self-Decoupled Magnetic Tactile Sensor},
  author={Dai, Huangzhe and Wu, Zheyan and Meng, Chenxian and Zhang, Chengqian and Zhao, Peng},
  journal={Magnetochemistry},
  volume={10},
  number={1},
  pages={6},
  year={2024},
  publisher={MDPI}
}

@misc{SupMat,
  note = "See supplementary material at  [URL inserted by editor] for details on complementary results from OOMMF simulations."
}

@misc{tetmagHertel,
author = {Hertel, Riccardo},
title = {tetmag},
url = {https://github.com/R-Hertel/tetmag}
}

\end{document}